\documentclass[twocolumn]{aastex701}

\usepackage{enumitem}
\usepackage{caption}

\begin{document}

\title{Ultraviolet and Optical Signatures of Late-time Disk Instabilities in Tidal Disruption Events}

\author[0000-0002-6347-3089]{Daichi Tsuna}
\affiliation{Center for Astrophysics $|$ Harvard \& Smithsonian, 60 Garden St, Cambridge, MA 02138, USA}
\email[show]{daichi.tsuna@cfa.harvard.edu}

\author[0000-0002-5814-4061]{V. Ashley Villar}
\affiliation{Center for Astrophysics $|$ Harvard \& Smithsonian, 60 Garden St, Cambridge, MA 02138, USA}
\affiliation{The NSF AI Institute for Artificial Intelligence and Fundamental Interactions}
\email{ashleyvillar@cfa.harvard.edu} 

\author[0000-0001-6806-0673]{Anthony L. Piro}
\affiliation{The Observatories of the Carnegie Institution for Science, Pasadena, CA 91101, USA}
\email{piro@carnegiescience.edu}

\author[0000-0003-2872-5153]{Samantha C. Wu}
\affiliation{The Observatories of the Carnegie Institution for Science, Pasadena, CA 91101, USA}
\affiliation{Center for Interdisciplinary Exploration \& Research in Astrophysics (CIERA), Physics \& Astronomy, Northwestern University, Evanston, IL 60202, USA}
\email{swu@carnegiescience.edu}  

\begin{abstract}
Tidal disruption events (TDEs) are unique probes of evolving accretion in supermassive black holes. Recent models of TDE disks show that they undergo brief thermal instabilities with temporal super-Eddington accretion at late times, which has been suggested as a possibility to explain the ubiquitous late radio emergence in TDEs. We model the ultraviolet (UV) and optical signatures of such disk instabilities, expected from the accretion power being reprocessed by the optically-thick outflow following super-Eddington accretion. Our model predicts brief UV-bright transients lasting for days, with luminosities of $10^{42}$--$10^{43}$ erg s$^{-1}$ in near-UV and $10^{41}$--$10^{42}$ erg s$^{-1}$ in optical for a typical TDE by a $10^6~M_\odot$ black hole. These could be detectable by near-future surveys such as ULTRASAT, Vera C. Rubin Observatory and Argus Array, for TDEs of redshifts out to $\approx 0.1$. We further conduct a search for these transients in existing nearby TDEs using data from the Zwicky Transient Facility, placing upper limits on the flare rate for each TDE of $1$--$2$ yr$^{-1}$ dependent on the outflow mass. In the era of future surveys, combined UV/optical and radio monitoring would be an important test to the disk instability phenomena, as well as its explanation for the late-time radio emission in TDEs.
\end{abstract}

\section{Introduction}
A star is tidally disrupted when it reaches too close to a supermassive black hole (BH), such that tidal forces overwhelm the star's self-gravity \citep{Hills75, Rees88, Phinney89}. These tidal disruption events (TDEs), observed in wavelengths spanning from radio to X-rays, are valuable probes for studying the physics of BH accretion and the environments surrounding supermassive BHs \citep[e.g.][for reviews]{Alexander20,Gezari21,Saxton21,Mockler25}.

About half of the disrupted stellar debris is gravitationally bound to the BH and eventually settles to form an accretion disk once its energy and angular momentum are significantly dissipated. The ultraviolet (UV) plateau observed in a number of TDEs indicates disk formation within a timescale of years after optical peak \citep{vanVelzen19,2024MNRAS.527.2452M,Guolo25}. 

Meanwhile, radio observations find late time brightening hundreds to thousands of days after optical peak \citep{Horesh21,Cendes22,Cendes24,Sfaradi24,Golay25} for a large fraction of TDEs (40\% of optical TDEs; \citealt{Cendes24}). Equipartition analysis of a large sample of TDEs favor the late radio emission in many of them being powered by a delayed launch of mildly relativistic ($\sim 0.1c$) outflows \citep{Cendes24,Goodwin25}. The origin(s) for such delayed outflows is unclear, while recent studies indicate connections with the evolution of the accretion onto the supermassive BH \citep{Alexander25}.

One possibility to explain the delayed outflow is that the disk formation itself is delayed. Such delayed circularization and disk formation relative to optical peak is argued in recent simulations, while the detailed timescale is still in active debate \citep[e.g.,][]{Ryu23,Price24,Steinberg24,Huang24,Huang25}\footnote{There are also scenarios that explain the delayed emergence by initially misaligned precessing jets \citep{Tchekhovskoy14,Teboul23,Lu24}, though it is unclear whether the high occurrence rates and low measured velocities \citep{Cendes24} can be reproduced in such scenarios.}. Another possibility is thermal instabilities in TDE disks occurring at late times \citep[][but see also \citealt{Alush25}]{Lu22,Piro25,Guo25}. Under these instabilities the disk can undergo brief, enhanced accretion onto the BH, which often exceeds the Eddington limit and could lead to mildly relativistic outflows \citep[e.g.,][]{Jiang24}. Recently, \cite{Wu25} find from detailed radio modeling that the sub-relativistic outflows expected from disk instabilities are a viable possibility to explain many of the events \citep[see also][]{Sato25}.

As various scenarios of delayed radio emission exist, a key question is whether there could be other unique predictions from such disk instabilities. In this work we consider the emission from the outflow at the onset of such disk accretion flares. We find that the flares are typically accompanied by brief UV-bright transients, lasting for a few days with peak luminosities of $10^{42}$--$10^{43}$ erg s$^{-1}$ in near-UV, making this scenario testable in high-cadence UV and optical surveys. 

Our work is constructed as follows. In Section \ref{sec:model} we present our model for calculating the UV/optical emission from the accretion flares, and the key features of the light curve evolution. We present a parameter study of this in Section \ref{sec:prospect}, and discuss the prospects of UV/optical surveys to test the existence of disk instabilities using these transients. In Section \ref{sec:search} we present our search for these transients in archival TDEs, using existing high-cadence optical data. We conclude in Section \ref{sec:conclusion}.

\section{Emission from Disk Outflows by the Accretion Flare}
\label{sec:model}
Our model for the accretion flare is inspired by time-dependent one-zone modeling of \cite{Piro25}. We have three parameters, the total mass ejected by the accretion flare $M_{\rm fl}$, the flare's duration $t_{\rm fl}$, and the characteristic disk radius $r_d$ during the flare. \cite{Piro25} find that the flare's duration is nearly independent of both the BH's mass $M_{\rm BH}$ and time from disruption, with $t_{\rm fl}\approx 1$--$2$ days. This arises from the disk radius scaling with the tidal radius 
\begin{eqnarray}
    r_T\approx \left(\frac{M_{\rm BH}}{M_*}\right)^{\!\!1/3}R_*\sim 100~R_\odot\left(\frac{M_{\rm BH}}{10^6~M_\odot}\right)^{\!\!1/3}\left(\frac{M_*}{M_\odot}\right)^{\!\!7/15},
    \label{eq:r_tidal}
\end{eqnarray}
by a nearly constant factor ($r_d\sim 5r_T$ in their model at the flaring state), and from $t_{\rm fl}$ being roughly the viscous time
\begin{eqnarray}
    t_{\rm visc}&\approx& \frac{1}{\alpha}\sqrt{\frac{r_d^3}{GM_{\rm BH}}} \sim 2~{\rm day}\left(\frac{r_d}{5r_T}\right)^{\!\!3/2}\left(\frac{M_*}{M_\odot}\right)^{\!\!7/10},
    \label{eq:t_visc}
\end{eqnarray}
where $\alpha\approx 0.1$ is the viscosity parameter. In equation (\ref{eq:r_tidal}) we have used the mass-radius relation of main sequence stars $R_*/R_\odot\approx(M_*/M_\odot)^{4/5}$. One sees that the dependence on $M_{\rm BH}$ cancels out when $r_d/r_T$ is fixed.

On the other hand, the value of $M_{\rm fl}$ is less certain, as it depends on the accretion history (as well as $\alpha$ and $M_{\rm BH}$) which is less clear especially at early times. In this work we take an empirical approach and consider masses of $M_{\rm fl}=10^{-2}$--$10^{-1}~M_\odot$, motivated by the results of \cite{Wu25} for explaining both the timescale and luminosity of the delayed radio detections in TDEs.

In this high accretion state, most of the accreted mass is ejected as a wind due to temporal super-Eddington accretion with rate
\begin{eqnarray}
    \dot{M}_w \approx \frac{M_{\rm fl}}{t_{\rm fl}} \sim 4~M_\odot\ {\rm yr}^{-1}\left(\frac{M_{\rm fl}}{10^{-2}M_\odot}\right)\left(\frac{t_{\rm fl}}{1\ {\rm day}}\right)^{-1},
\end{eqnarray}
which highly exceeds the Eddington-limited accretion rate $\dot{M}_{\rm Edd}\sim 10^{-2}M_\odot\ {\rm yr}^{-1}(M_{\rm BH}/10^6M_\odot)$.

We assume the wind is launched quasi-spherically from the disk due to accretion at a characteristic radius $r_{\rm in}$ with constant rate $\dot{M}\approx \dot{M}_w \approx M_{\rm fl}/t_{\rm fl}$, carrying an accretion power of \footnote{For our expected case of $r_{\rm in}\sim 10^2$--$10^3R_g$, this formulation results in an energy efficiency $L_{\rm acc}/\dot{M}_wc^2$ roughly consistent with the  efficiency of $\sim 0.1$--$1~\%$ obtained in numerical simulations of super-Eddington disks \citep[e.g.,][]{Jiang19,Yoshioka24,Zhang25_2}.}
\begin{eqnarray}
    L_{\rm acc}(t \leq t_{\rm fl}) &\approx& \frac{GM_{\rm BH}\dot{M}_w}{r_{\rm in}} \nonumber \\
    &\sim& 2\times 10^{44}\ {\rm erg\ s^{-1}}\left(\frac{\dot{M}_w}{M_\odot\ {\rm yr}^{-1}}\right)\left(\frac{r_{\rm in}}{300R_g}\right)^{\!\!-1}.
    \label{eq:L_acc}
\end{eqnarray}
Here $R_g=GM_{\rm BH}/c^2\sim 1.5\times 10^{11}$ cm $(M_{\rm BH}/10^6M_\odot)$ is the BH's gravitational radius.
The launching radius $r_{\rm in}$ could correspond to the ``spherization radius" \citep{Shakura73} inside which the local accretion power exceeds the Eddington limit $L_{\rm Edd}=4\pi GM_{\rm BH}c/\kappa$,
\begin{equation}
    r_{\rm Sph}\approx \frac{\kappa \dot{M}_w}{4\pi c} \sim 5\times 10^{13}\ {\rm cm}\left(\frac{\kappa}{0.3\ {\rm cm^2\ g^{-1}}}\right)\left(\frac{\dot{M}_w}{M_\odot\ {\rm yr}^{-1}}\right),
\end{equation}
or $r_{\rm in}\approx r_d$ if $r_d<r_{\rm Sph}$ and the accretion is super-Eddington everywhere in the disk. We therefore adopt 
\begin{eqnarray}
    r_{\rm in} = {\rm min}(r_d, r_{\rm Sph}).
\end{eqnarray}
The opacity $\kappa$ is expected to be dominated by scattering opacity for the density and temperature of our interest, and we adopt a constant $\kappa \approx 0.3\ {\rm cm^2\ g^{-1}}$ for a solar-like composition.

For the typical case of a $10^6$--$10^7M_\odot$ BH disrupting a low-mass star of $M_*\lesssim M_\odot$, we expect the relation $r_d\approx 5r_T<r_{\rm Sph}$, and thus $r_{\rm in}\approx r_d$. Taking $r_d\approx 5r_T$, we can obtain the total energy budget from accretion power as
\begin{eqnarray}
    L_{\rm acc}t_{\rm fl} &\approx& \frac{GM_{\rm BH}M_{\rm fl}}{r_d}\nonumber \\
    &\sim& 8\times 10^{49}\ {\rm erg}\nonumber\\
    &&\times \left(\frac{M_{\rm fl}}{10^{-2}M_\odot}\right) \left(\frac{M_{\rm BH}}{10^6~M_\odot}\right)^{\!\!2/3}\left(\frac{M_*}{M_\odot}\right)^{\!\!-7/15}.
\end{eqnarray}
This is consistent with the energy for the radio analysis adopted in \cite{Wu25} of $1.4\times 10^{50}\ {\rm erg}~(M_{\rm fl}/10^{-2}~M_\odot)$, within a factor of a few across a BH mass range of $10^6$--$10^7~M_\odot$\footnote{We note that our method for calculating this energy slightly differs from that of \cite{Wu25}. \cite{Wu25} fix the outflow’s velocity structure without their dependence on the BH mass, while we include the mass dependence via equation (\ref{eq:L_acc}).}.

At $r_{\rm in}$ we expect the accretion power to be shared nearly equally between internal and kinetic energy, but the wind is highly optically thick. Due to adiabatic expansion, internal energy is degraded as it is converted to kinetic energy via $PdV$ work, and the observed luminosity is generally reduced from $L_{\rm acc}$. 

Our aim is to semianalytically calculate the emission from such time-dependent wind outflows. We follow the spirit of \cite{Piro20} (see their Figure 1 for context), but we build the model slightly differently so that it can properly solve under abrupt changes in the accretion power and outflow like considered here.

\subsection{Light curve}
Our approach below is to split the wind into multiple ``shells" launched at times $\tilde{t}$ ($0<\tilde{t}<t_{\rm fl}$) and calculate the radiative output of each shell at time $t'$ from launch, to obtain the luminosity observed at time $t=\tilde{t}+t'$.

\begin{figure*}
    \centering
\includegraphics[width=0.9\linewidth]{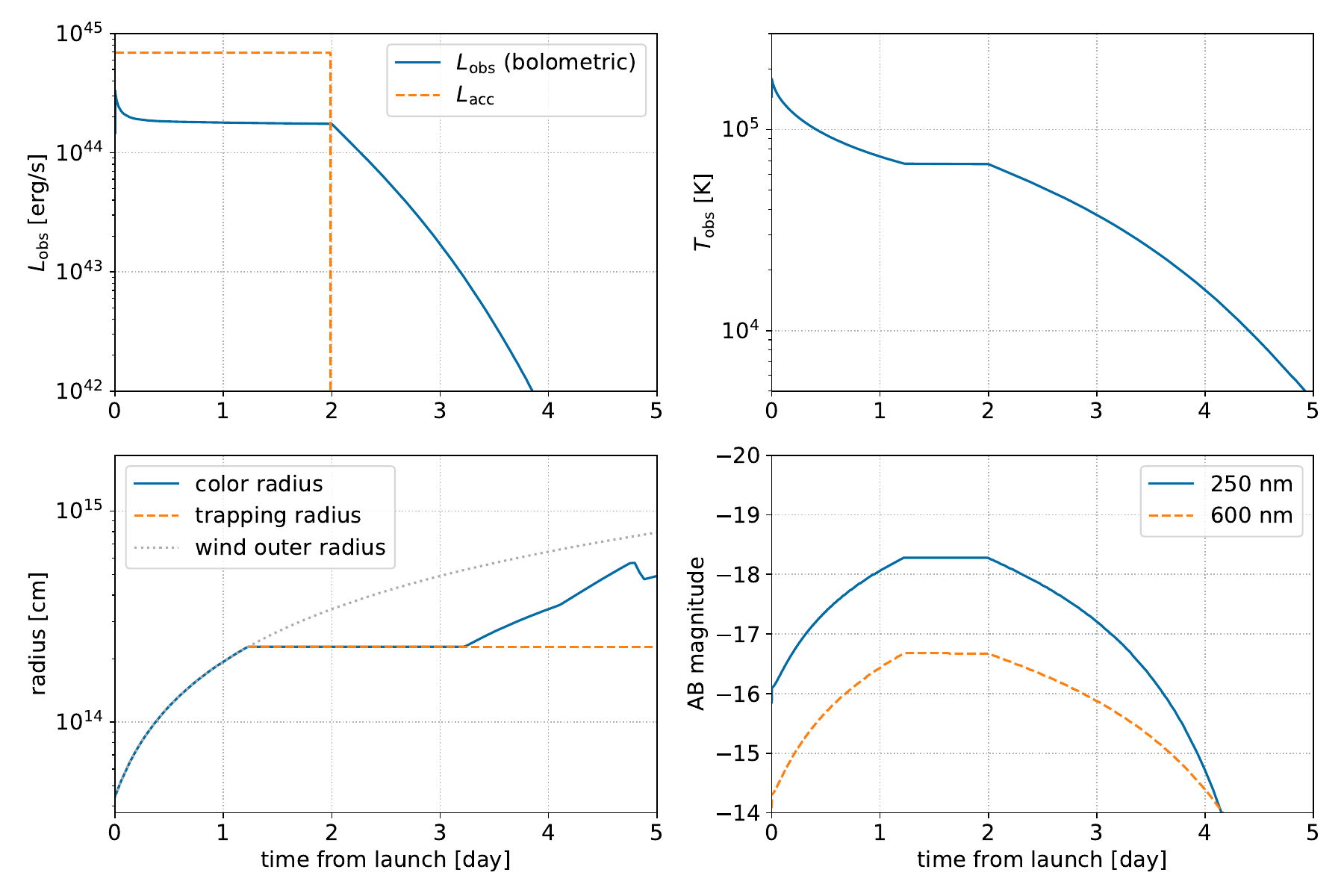}
    \caption{Light curve predictions for parameters: $M_{\rm BH}=10^6M_\odot$, $r_d=300R_g$, $M_{\rm fl}=0.02~M_\odot$, and $t_{\rm fl}=2$ days. Top panels show the bolometric luminosity and emission temperature, and the bottom right panel shows the AB magnitudes in the near-UV and optical under a greybody approximation. Bottom left panel shows the evolution of various radii defined in Section \ref{sec:temperature}.}
    \label{fig:single_model}
\end{figure*}

Let us consider a shell of wind material launched over a time $\Delta t$ at radius $r_{\rm in}$ and time $\tilde{t}$, with internal energy
\begin{eqnarray}
    E_{\rm int, 0} \approx \frac{1}{2}L_{\rm acc} \Delta t
\end{eqnarray}
and with velocity
\begin{eqnarray}
v_w\sim \sqrt{(L_{\rm acc}/2)/(\dot{M}_w/2)}\sim 0.06c\left(\frac{r_{\rm in}}{300R_g}\right)^{-1/2}.
\label{eq:v_w}
\end{eqnarray}
At a time $t'$ from launch, the shell increases its volume as $V_{\rm sh}=4\pi[r_{\rm in}+(v_w t')]^2(v_w\Delta t)$, and its internal energy $E_{\rm int}$ (assumed to be radiation dominated) evolves as
\begin{eqnarray}
    \frac{dE_{\rm int}}{dt'} &=& -\frac{E_{\rm int}}{3V_{\rm sh}}\frac{dV_{\rm sh}}{dt'} - L_{\rm rad} \nonumber \\
    &=& -\frac{2E_{\rm int}}{3(t'+r_{\rm in}/v_w)} - \frac{E_{\rm int}}{t_{\rm diff}},
\end{eqnarray}
where the first term is adiabatic cooling by $PdV$ work, and the second term is the radiative cooling regulated by the diffusion of photons through the wind over a timescale \citep{Piro20}
\begin{eqnarray}
    t_{\rm diff} &\approx& \frac{\kappa \dot{M}_w}{4\pi v_wc} \times \left[\frac{r_w - (r_{\rm in}+v_wt')}{r_w}\right]^2 \nonumber \\
    &\approx& \frac{\kappa \dot{M}_w}{4\pi v_wc} \times \left[\frac{v_w\tilde{t}}{r_{\rm in}+v_w(\tilde{t}+t')}\right]^2
    \label{eq:t_diff}
\end{eqnarray}
Here $r_w=r_{\rm in}+v_w(\tilde{t}+t')$ is the outer edge of the wind. The differential equation in $t'$ has an analytical solution, with initial condition of $E_{\rm int}=E_{\rm int,0}$ at $t'=0$, of
\begin{eqnarray}
    E_{\rm int}(t') &=&  E_{\rm int, 0}\left(\frac{r_{\rm in}}{r_{\rm in}+v_w t'}\right)^{2/3} \nonumber \\
    &\times&\exp\left\{-\frac{B^3}{3A}\left[\left(1+\frac{1}{B}\frac{t'}{\tilde{t}}\right)^3-1\right]\right\}
\end{eqnarray}
where $A,B$ are dimensionless parameters that are independent of $t'$,
\begin{eqnarray}
    A= \frac{1}{\tilde{t}}\frac{\kappa \dot{M}_w}{4\pi v_w c}, \ B=1+\frac{r_{\rm in}}{v_w\tilde{t}}\ .
\end{eqnarray}
The corresponding luminosity $L_{\rm rad}=E_{\rm int}/t_{\rm diff}$ is
\begin{eqnarray}
    L_{\rm rad}(t') &=& \frac{L_{\rm acc}\Delta t}{2A\tilde{t}}\left(B+\frac{t'}{\tilde{t}}\right)^2 \left(\frac{r_{\rm in}}{r_{\rm in}+v_w t'}\right)^{2/3} \nonumber \\
    &\times&\exp\left\{-\frac{B^3}{3A}\left[\left(1+\frac{1}{B}\frac{t'}{\tilde{t}}\right)^3-1\right]\right\}.
    \label{eq:L_rad_shell}
\end{eqnarray}

We can now integrate over all the emitted shells to obtain the light curve as a function of $t$. Using the solution for $L_{\rm rad}(t')$ and substituting $t'=t-\tilde{t}$ for each wind shell launched at $\tilde{t}$, we sum up the contributions from all shells emitted within $t$ and obtain
\begin{eqnarray}
    L_{\rm obs}(t)  &=& \int_0^t d\tilde{t}~ \frac{L_{\rm acc}(\tilde{t})}{2A\tilde{t}}\left(B+\frac{t-\tilde{t}}{\tilde{t}}\right)^{\!\!2} \left[\frac{r_{\rm in}}{r_{\rm in}+v_w (t-\tilde{t})}\right]^{\!2/3} \nonumber \\
    &\times&\exp\left\{-\frac{B^3}{3A}\left[\left(1+\frac{1}{B}\frac{t-\tilde{t}}{\tilde{t}}\right)^3-1\right]\right\},
    \label{eq:L_rad}
\end{eqnarray}
The top-left panel of Figure \ref{fig:single_model} shows a light curve for our fiducial parameter set of $M_{\rm BH}=10^6M_\odot$, $M_{\rm fl}=0.02~M_\odot$, $r_d=300R_g$, and $t_{\rm fl}=2$ days. After a brief decline during $t\ll r_{\rm in}/v_w\sim 0.3$ days as the diffusion time evolves \citep{Piro20}, the light curve settles to a plateau-like morphology. The plateau is partly due to our assumed top-hat time-dependence of $\dot{M}$ and $L_{\rm acc}$. The observed luminosity at this phase is a factor of several lower than the input luminosity $L_{\rm acc}$ due to the adiabatic loss. Once the outflow shuts off at $t=t_{\rm fl}$, the light curve drops sharply over a timescale of a few days, which is set by the diffusion timescale in the wind.

\subsection{Temperature Evolution}
\label{sec:temperature}
We calculate the temperature of the emission by approximately solving the reprocessing by the wind. Following \cite{Piro20}, we obtain this by calculating the color radius $r_c$, the outermost radius where thermalization occurs.

At a given time $t$, the density profile of the wind varies during and after the flare as
\begin{eqnarray}
    \rho(r,t<t_{\rm fl}) &=&  \frac{\dot{M}_w}{4\pi r^2 v_w}~ (r < r_w=r_{\rm in}+v_w t),\\
    \rho(r,t>t_{\rm fl}) &=& 
    \left\{\begin{array}{@{}l@{\quad}l}
       \frac{\dot{M}_w}{4\pi r^2 v_w} & (r_{\rm in}+v_w(t-t_{\rm fl}) < r < r_w)\\
      0 & ({\rm otherwise}).
    \end{array}
    \right.
\end{eqnarray}

The diffusion time in equation (\ref{eq:t_diff}) indicates that the bulk of the radiation is emitted around when the wind expandse to the trapping radius
\begin{eqnarray}
    r_{\rm tr} \approx  r_{\rm in} + \frac{\kappa \dot{M}_w}{4\pi c}.
    \label{eq:r_trap}
\end{eqnarray}
At the earliest stage when $r_w<r_{\rm tr}$, photons travel by advection throughout the wind, and we adopt the temperature set by adiabatic cooling out to the outermost radius of the wind, with $r_c\approx r_w$ and
\begin{eqnarray}
    T_{\rm obs} \approx \left(\frac{L_{\rm obs}}{4\pi r_w^2 v_w a}\right)^{1/4}.
    \label{eq:T_obs_advection}
\end{eqnarray}
Once $r_w > r_{\rm tr}$, above $r_{\rm tr}$ the luminosity is roughly constant, and the temperature could be well-represented by the diffusion approximation as \citep{Piro20}
\begin{eqnarray}
    L_{\rm obs} = - \frac{4\pi r^2 ac}{3\kappa \rho}\frac{\partial T^4}{\partial r} .
\end{eqnarray}
This leads to a temperature profile
\begin{eqnarray}
    T(r,t)^4 \approx T_{\rm out}^4+\int_r^{r_w}\frac{3\kappa\rho L_{\rm obs}}{4\pi r^2 ac},
\end{eqnarray}
where $T_{\rm out}$ is the temperature at the outer edge, which we solve for by fixing the temperature at $r_{\rm tr}$ as
\begin{eqnarray}
    T(r=r_{\rm tr},t)  \approx \left(\frac{L_{\rm obs}}{4\pi r_{\rm tr}^2 v_w a}\right)^{1/4}.
    \label{eq:T_trap}
\end{eqnarray}
In the density regimes of interest the opacity is scattering-dominated, and the color radius $r_c(t)$ that determines the temperature is interior to the photosphere. This radius is solved by
\begin{eqnarray}
    \int_{r_c}^{r_w}\sqrt{3\kappa_{\rm abs}\kappa}~\rho dr = 1
    \label{r_color}
\end{eqnarray}
For $\kappa_{\rm abs}$, we adopt a Kramer's opacity law motivated for bound-free and free-free absorption, with a sharp cutoff at low temperatures due to hydrogen recombination
\begin{eqnarray}
    \kappa_{\rm abs}(\rho, T) = 
    \left\{\begin{array}{@{}l@{\quad}l}
       C\rho T^{-3.5} & (T\geq T_{\rm rec})\\
       C\rho T_{\rm rec}^{-3.5}(T/T_{\rm rec})^\beta & (T<T_{\rm rec}),
    \end{array}
    \right.
\end{eqnarray}
where the parameters are in cgs units, and $C$, $T_{\rm rec}$ are dependent on the composition of the wind. Here we set $\beta=15$ to mimic the sharp cutoff at $T<T_{\rm rec}$, and adopt constant values ($C$, $T_{\rm rec}$)$=$($4\times 10^{25}$ cgs, $6000$ K) motivated for solar abundance gas \citep[e.g. Appendix of][]{Matsumoto21}. The detailed choice of $\beta$ does not affect our conclusions, as we later find emission temperatures to be much larger than $T_{\rm rec}$.

When $r_c$ solved by this procedure satisfies $r_c>r_{\rm tr}$ (``thermalization-dominated temperature" in \citealt{Piro20}), $T_{\rm obs}$ would be the temperature at the color radius
\begin{eqnarray}
    T_{\rm obs} \approx \left[T_{\rm out}^4+\int_{r_c}^{r_w}\frac{3\kappa\rho L_{\rm obs}}{4\pi r^2 ac}\right]^{1/4}\ (r_c > r_{\rm tr}).
\end{eqnarray}
On the other hand if $r_c<r_{\rm tr}$ (``trapping-dominated temperature" in \citealt{Piro20}), the temperature of the emission is set by the trapping radius. We adopt the temperature at $r_{\rm tr}$ of
\begin{eqnarray}
     T_{\rm obs} \approx \left(\frac{L_{\rm obs}}{4\pi r_{\rm tr}^2 v_w a}\right)^{1/4}\ (r_c < r_{\rm tr}),
     \label{eq:T_obs_trapping}
\end{eqnarray}
and update $r_c=r_{\rm tr}$. 

\begin{figure*}
    \centering
    \includegraphics[width=\linewidth]{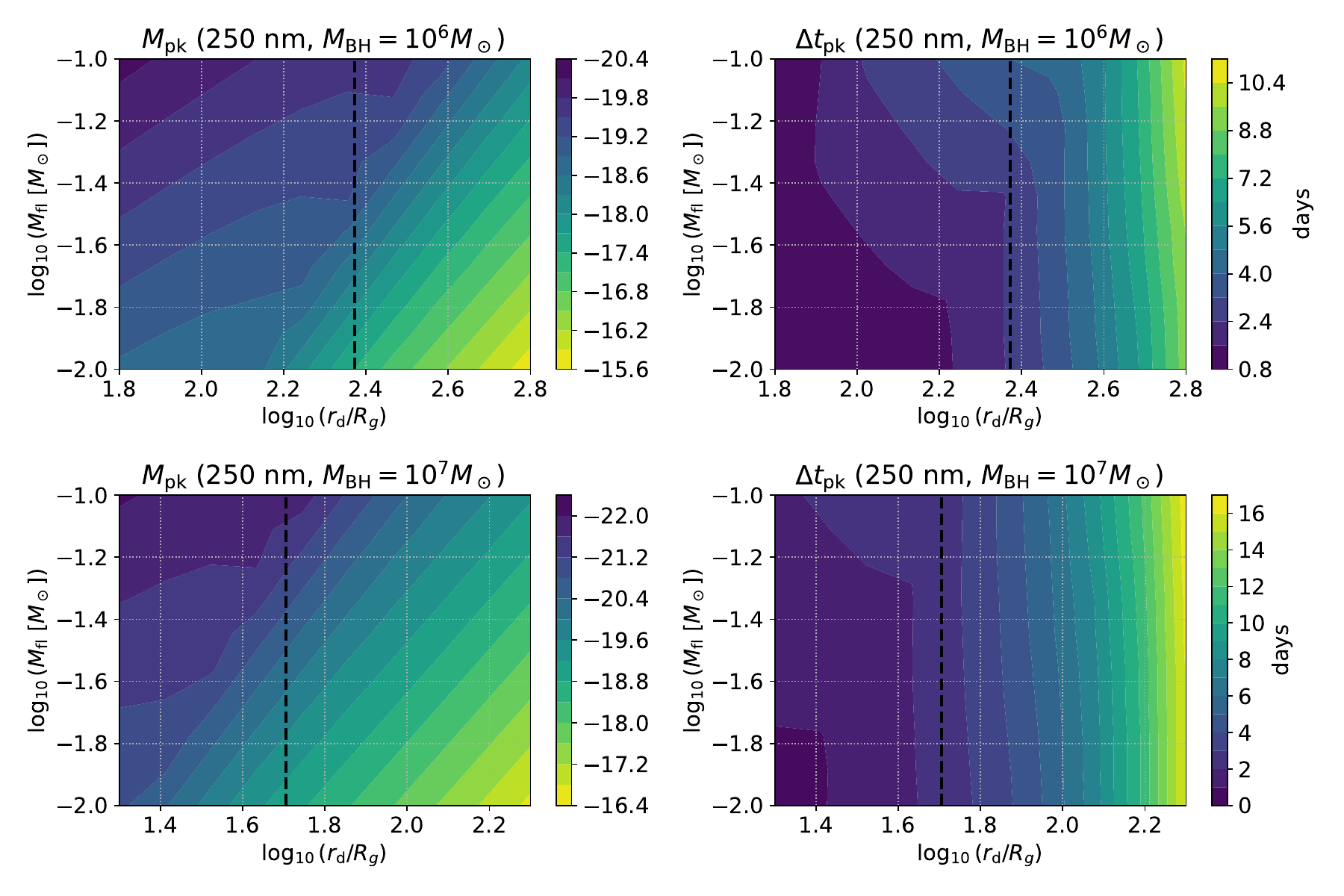}
    \caption{Emission properties in the near-UV (250 nm). Left and right panels respectively show the peak magnitude and duration within 1 mag from peak. Top and bottom panels are respectively for $M_{\rm BH}=10^6, 10^7~M_\odot$. Vertical dashed lines show the radius $r_d\approx 5r_T$ inspired from the model of \cite{Piro25}, where $r_T$ is the tidal radius (see main text).}
    \label{fig:param_study_UV}
\end{figure*}

The temperature predictions solved this way are shown in the top-right panel of Figure \ref{fig:single_model}. Initially photons are completely advected by the wind, and the temperature drops as given in equation (\ref{eq:T_obs_advection}). Then it enters the trapping-dominated temperature where $T_{\rm obs}$ is set by equation (\ref{eq:T_obs_trapping}). Here $T_{\rm obs}$ plateaus at $\approx 1$--$2$ days because both $L_{\rm obs}$ and $r_{\rm tr}$ are nearly constants. The plateau only appears at sufficiently low values of $M_{\rm fl}$ when $r_{\rm tr}/v_w < t_{\rm fl}$ is satisfied. For much larger $M_{\rm fl}$ (and $r_{\rm tr}$), the trapping-dominated regime applies only at $t>t_{\rm fl}$, and $T_{\rm obs}$ then monotonically declines with time because $L_{\rm obs}(t>t_{\rm fl})$ drops with time.

Since $\kappa_{\rm obs}$ is strongly temperature-dependent as $\kappa_{\rm abs}\propto T^{-3.5}$, the thermalization-dominated regime (where $r_c>r_{\rm tr}$) is reached only when the temperature at the outer edge of the wind becomes low enough, about $10^4$ K for this parameter set. The kink in the color radius evolution after this reflects the strong turnover of $\kappa_{\rm abs}$ around $T=T_{\rm rec}$.

In the bottom-right panel of Figure \ref{fig:single_model} we show predictions for the AB magnitude in the near-UV (250 nm) and optical (600 nm), assuming the transient has greybody spectra with temperature $T_{\rm obs}$ and luminosity $L_{\rm obs}$. Due to the high temperature of $\sim 10^5$ K at early times, the AB magnitudes peak closer to the end of the flare. For this fiducial case, the magnitudes in near-UV and optical are respectively $<-17$ mag and $<-15$ mag over a few days. The top-hat feature of the light curve in these bands is again due to our adopted time dependence for $\dot{M}$ and $L_{\rm acc}$, and could more smoothly evolve in reality.

Here we comment on two simplifications of our model that may be improved in future studies. Our formalism assumes spherical symmetry, with our temperature estimates corresponding to those of emission reprocessed by the optically thick disk wind. Super-Eddington outflows are expected to have angular dependence, with lower density and optical depth in polar regions. While the observability of the incident disk emission (likely in X-rays) is thus strongly viewing-angle dependent, the reprocessed emission (in UV/optical) that undergo repeated scattering is nearly isotropic and largely insensitive to the viewing angle. This has also been found in detailed multi-dimensional radiation transfer models of super-Eddington outflows \citep[e.g.,][]{Parkinson25}. Our model assumes full reprocessing with covering fraction of unity, and in reality we expect the incident emission can avoid reprocessing in the polar direction. We nonetheless expect this to be an order-unity correction for our UV/optical emission from geometrically thick super-Eddington flows.

Furthermore, we have made predictions for the UV/optical emission under greybody spectra. In the high temperatures of our interest, the emergent spectrum is expected to be characterized by a greybody with a deviation in the infrared/optical, due to the large and frequency-dependent opacity of free-free (and bound-free) transitions at those long wavelengths \citep[e.g.,][]{Lu20,Chen25}. A detailed spectral modeling has been explored in \cite{Chen25} in the context of fast blue optical transients, which share similar parameter spaces with ours in outflow rate and temperature. From their Section 3.1, we expect the spectra to be nearly a greybody out to a characteristic break wavelength (their equation 24 and Figure 3)
\begin{eqnarray}
    \lambda_b &\approx& 1.4\ {\rm \mu m}\left(\frac{\dot{M}_w}{M_\odot{\rm yr}^{-1}}\right)^{\!1/8}
    \left(\frac{v_w}{0.1c}\right)^{\!21/16}\left(\frac{L_{\rm obs}}{10^{44}{\rm erg\ s^{-1}}}\right)^{\!3/16},
\end{eqnarray}
where we use equation (\ref{eq:T_trap}) and $r_{\rm tr}\gg r_{\rm in}$ to obtain the scaling with $\dot{M}_w$. While more detailed wavelength-dependent calculations are valuable, we expect our UV/optical predictions to be reasonable approximations for the purpose of our subsequent discussions regarding parameter exploration and observational prospects.

\section{Parameter Exploration}
\label{sec:prospect}

\begin{figure*}
    \centering
    \includegraphics[width=\linewidth]{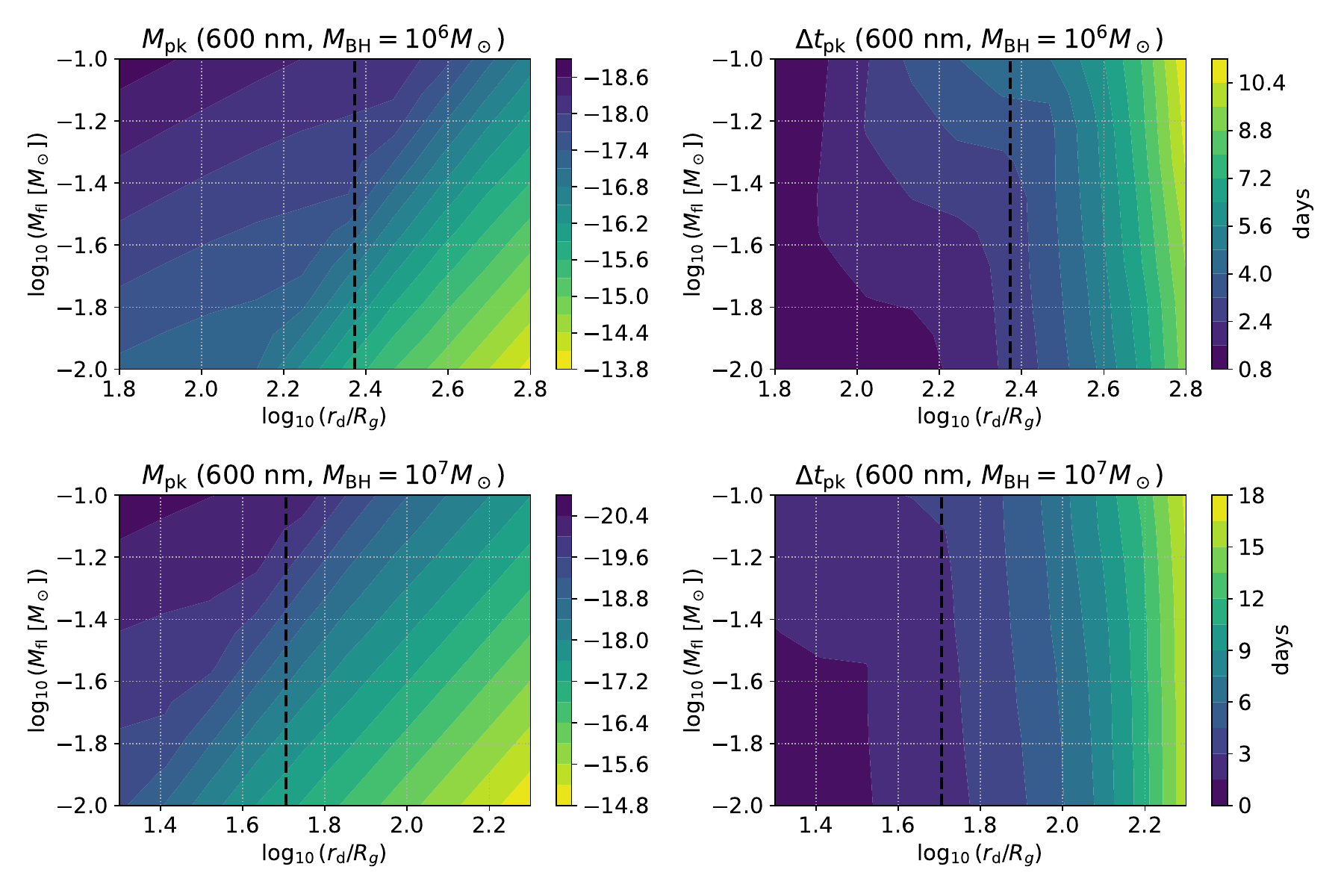}
    \caption{Same as Figure \ref{fig:param_study_optical}, but for the optical (600 nm).}
    \label{fig:param_study_optical}
\end{figure*}

The most promising instruments to detect these transients are surveys with short cadences around a few days or less, such as the Zwicky Transient Facility \citep[ZTF;][]{Bellm19}, ULTRASAT \citep{Shvartzvald24} and Vera C. Rubin Observatory’s Legacy Survey of Space and Time \citep[LSST;][]{Ivezic19}. The detectability of these flares in optical and UV varies with both $r_d$ and $M_{\rm fl}$, as they control both the energetics and temperature of the flare. We calculate a grid of light curves varying these two parameters, and measure the peak magnitude $M_{\rm pk}$ and duration $\Delta t_{\rm pk}$ of the light curve within $1$ mag from peak.

We adopt $M_{\rm fl}=0.01$--$0.1~M_\odot$ favored from late time radio emission modeling \citep{Wu25}, and choose a range of $r_d$ around $r_d\approx 5r_T$ approximately found in the model of \cite{Piro25}. This value of $r_d$, corresponding to the disk's outer radius, is also roughly reproduced in multi-wavelength analyses of late-time plateau emission in TDEs \citep[][their Figure 7]{Guolo25}. We set the flare duration $t_{\rm fl}$ as the viscous time of the disk, scaling with $r_d$ as in equation \ref{eq:t_visc} (with $M_*=M_\odot$).

Figure \ref{fig:param_study_UV} shows the range of $M_{\rm pk}$ and $\Delta t_{\rm pk}$ at near-UV (250 nm) for the parameter space of $r_d$ and $M_{\rm fl}$, with $M_{\rm BH}$ of $10^6~M_\odot$ and $10^7~M_\odot$ in the top and bottom panels. For our fiducial choice of $r_d=5r_T$ (dashed lines) and $M_{\rm BH}=10^6~M_\odot$, we find the near-UV emission to peak at magnitudes between $-17$ and $-19$ mag, with durations of $3$--$4$ days. ULTRASAT, with a single-visit sensitivity of 22.5 mag at 250 nm and a cadence of 4 days \citep{Shvartzvald24}, could detect such late-time UV flares in a large fraction of TDEs out to redshift $z\approx 0.1$. Thus combined UV and radio monitoring would offer strong direct tests for the disk instability scenario, as radio measurements can infer (under the equipartition assumption) the radius evolution and hence the launch time of the super-Eddington outflow. The case of a $10^7~M_\odot$ BH predicts brighter flares for the same assumptions of $r_d=5r_T$, as the launch radius scaled by the BH's gravitational radius $r_{\rm in}/r_g$ is smaller (equation \ref{eq:L_acc}). However, this also leads to a faster wind velocity and hence a shorter duration ($\lesssim 3$ days).

Figure \ref{fig:param_study_optical} shows the same parameter grid as Figure \ref{fig:param_study_UV} but for the optical at 600 nm. For a typical supermassive BH mass of $M_{\rm BH}=10^6~M_\odot$ and $r_d\approx 5r_T$, the emission in the optical is $-14$ to $-17$ mag ($10^{41}$--$10^{42}$ erg s$^{-1}$) and lasts for $3$--$5$ days. This emission would be buried by the main TDE emission at early times but may be visible on year timescales after the optical peak. Such emission would be a good target in the Rubin era, for sources also out to $z\approx 0.1$. 

The disk size could vary from our fiducial choice of $r_d\approx 5r_T$, especially at early stages of debris circularization before further viscous evolution sets the disk size. For an orbital eccentricity and pericenter distance of the disrupted star of $e\approx 1$ and $r_p\approx r_T$, from angular momentum conservation the debris would settle at a circularization radius of $(1+e)r_p \approx 2r_T$. Thus at early times, $r_d$ could be up to a factor few smaller than our fiducial $r_d\approx 5r_T$, though the difference would be smaller as one probes later times of the TDE.\footnote{On the other hand, due to viscous spreading $r_d$ could become larger than our fiducial choice, at times much later than years after TDE of our interest. However the occurrence of the flare is predicted to be increasingly rarer at later times as the fallback rate drops, effectively shutting off within several years from TDE \citep{Piro25}.} A smaller $r_d$ would lead to a stronger accretion power (equation \ref{eq:L_acc}) with faster outflows (equation \ref{eq:v_w}), both leading to a brighter and faster transient. Inspecting Figures \ref{fig:param_study_UV} and \ref{fig:param_study_optical}, variations in $r_d$ by a factor of 2 changes the UV/optical luminosity by around $1$--$1.5$ mag and timescale by a factor $\sim 2$.

A confident detection of the flare may require detection over multiple epochs, which is generally challenging for the predicted short durations of the flares. The Argus Array \citep{Law22}, with deep coadded sensitivity for $\approx 23$ mag for a 1 day cadence, would be particularly powerful for this purpose. The LSST deep drilling field could also be similarly powerful \citep{Gris23}, for a small subset of TDEs that occur in the field.

\section{Search in Optical Light Curve Data of Existing Tidal Disruption Events}
\label{sec:search}

\begin{figure}
    \centering
    \includegraphics[width=\linewidth]{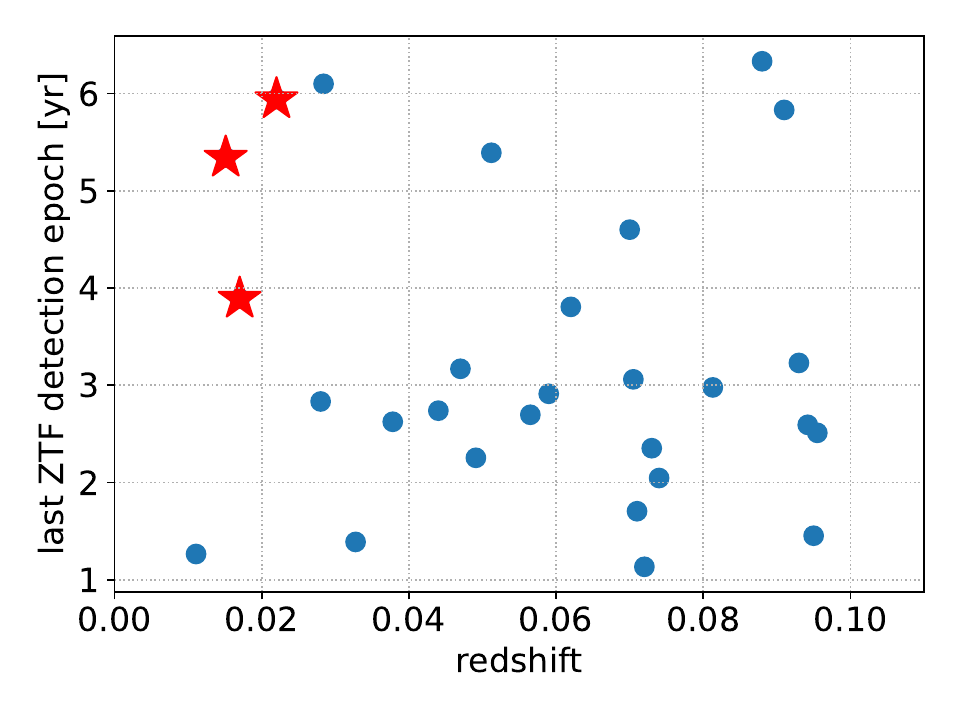}
    \caption{Redshift and epoch of last ZTF detection (relative to discovery), for the 28 TDE samples analyzed in this work. Stars show three TDEs used for constraints on the flare rate, AT 2019qiz, AT 2019azh, and AT 2021ehb (see main text).}
    \label{fig:z_vs_last_det}
\end{figure}

Given the flare's luminosity, we expect that the flares may even be detectable by existing optical surveys, especially for nearby TDEs that result in brighter flares. For example, TDEs at distances $\lesssim 100$ Mpc could have flares of $\lesssim 21$ mag in optical, potentially within reach of surveys like ZTF.

Motivated by this, we conduct a search of late-time flares in existing TDEs, using light curve data from the OTTER database \citep{Franz25}. We select a sample of 28 TDEs with (i) redshift less than $0.1$, (ii) host correction applied, and (iii) with at least one ZTF detection (flagged by $>3\sigma$ confidence in OTTER) at $>1$ year from discovery. We show the redshift and the epoch of the last ZTF detection (from discovery) in Figure \ref{fig:z_vs_last_det} (for details of the events see Appendix \ref{sec:TDE_samples}). About half of the samples have detections as late as 3 years after discovery, and four TDEs are at luminosity distances $D_L<100$ Mpc.

As TDEs generally have disk-dominated plateau emission at late times \citep[e.g.,][]{vanVelzen19}, we search for any short-duration excess that could be consistent with a flare on top of a longer plateau. For each TDE, we extract all host-subtracted ZTF data at $>1$ yr from discovery, and compare the flux of each detection with the surrounding flux measurements in the same band at $\pm$(3--50) days. Assuming the flux uncertainties are Gaussian, we estimate the significance as ${\rm min}[(F-f)/\sqrt{dF^2+df^2}]$, evaluated using all the surrounding flux measurements $f\pm df$ around the detection with flux $F\pm dF$. We exclude measurements within 3 days as they could be part of a single flare, and beyond 50 days to prevent multiple flares from blending into the analysis. Our search is hence less sensitive to flares frequently occurring with intervals $\lesssim 50$ days, as the flares can blend into both $F$ and $f$. However we believe these short intervals are unlikely at $\gtrsim 1$ yr after the TDE, based on existing disk models \citep{Lu22,Piro25}.

Our analysis did not find any strong candidates of a disk instability flare. The excess found with high significance were all second peaks of events claimed as repeating (partial) TDEs in the literature, AT 2020vdq \citep{Somalwar25} and AT 2022dbl \citep{Lin24, Hinkle24}. These have luminous peaks ($\nu L_\nu \sim 10^{43}$ erg s$^{-1}$) and slow declines ($\approx 1$ month) in the optical, inconsistent with the expectations from our model in Section \ref{sec:prospect}. Hence we believe these excesses are unlikely to be due to disk instabilities.

\begin{figure*}
    \centering
    \includegraphics[width=0.95\linewidth]{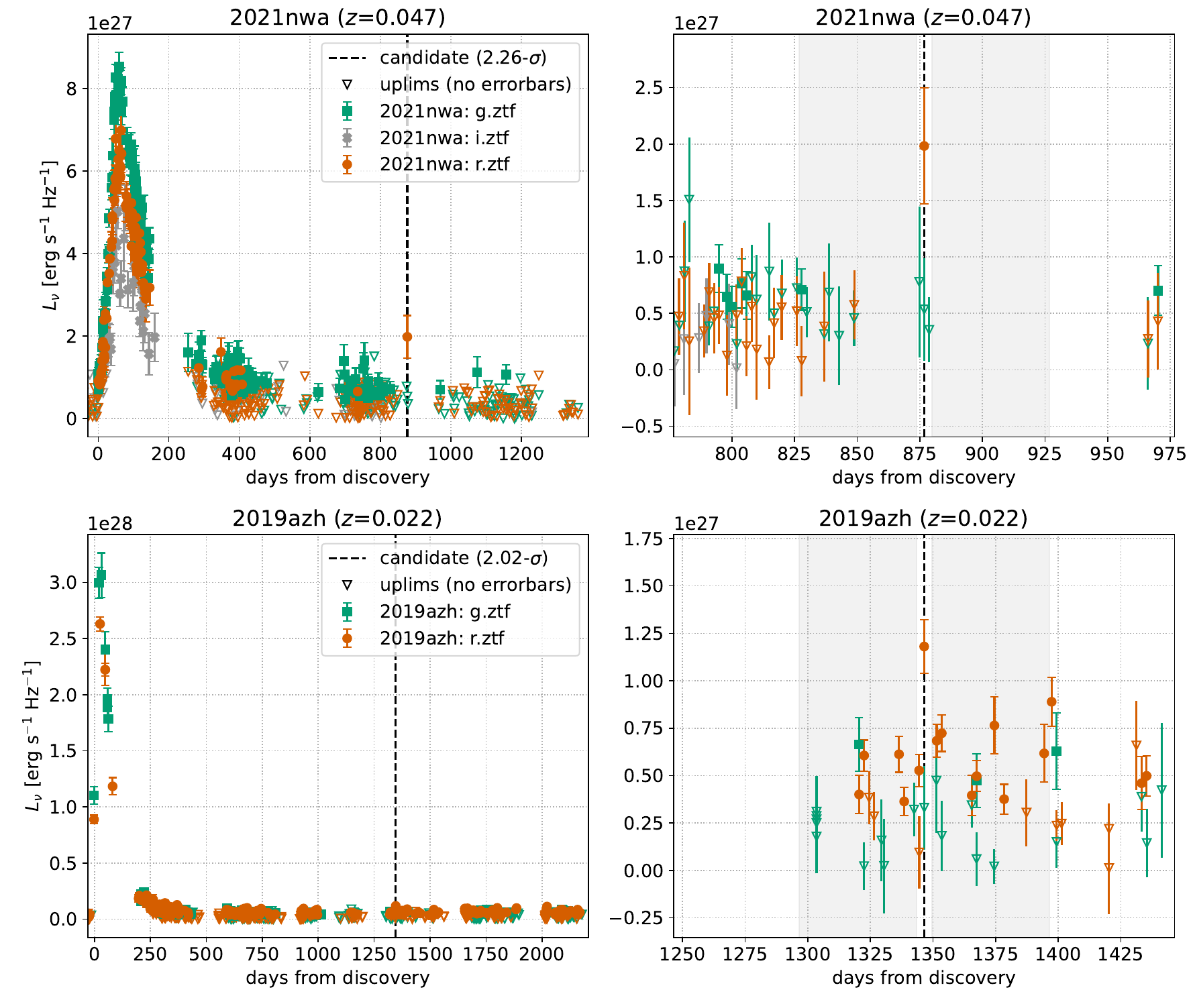}
    \caption{The two highest-significance candidates found in our analysis, from AT2019azh and AT2021nwa. Left panels show the full light curve, while the right panels show that at $\pm 100$ days around the candidate. Shaded regions show the temporal regions used in our search for the flux comparison.}
    \label{fig:candidates}
\end{figure*}

The remaining candidates are all low significance and not strong enough to claim a detection. In Figure \ref{fig:candidates} we show the two highest-significance candidates from AT 2021nwa and AT 2019azh, which are both flagged as an $r$-band excess. We believe these candidates are also unlikely to be disk instability flares, as the $g$-band flux in the vicinity of each candidate is not enhanced like the $r$-band flux.

We finally place rough rate constraints of these flares by artificial injections of our theoretical light curve models into ZTF data. For this purpose we focus our analysis on the three most constraining TDEs, which are nearby ($D_L< 100$ Mpc) and have long ZTF baselines of several years: AT 2019qiz, AT 2019azh, and AT 2021ehb. 

Into the observed data, we add signals of flux
\begin{eqnarray}
    f_\nu = \frac{1+z}{4\pi D_L^2}L_{\nu'}[L_{\rm obs},T_{\rm obs}],
\end{eqnarray}
where $\nu'=(1+z)\nu$ is the source-frame frequency and $L_{\nu'}$ is calculated assuming a blackbody (at source-frame) with bolometric luminosity $L_{\rm obs}$ and temperature $T_{\rm obs}$. The observer-frame frequency $\nu$ is represented with the central frequency of each filter, $6.34\times 10^{14}$ Hz and $4.72\times 10^{14}$ Hz for $g$ and $r$ bands, respectively. For the signal we again vary the flare mass as $M_{\rm fl}=0.01$--$0.1~M_\odot$, while we fix the disk radius as $r_d=5r_T$ and adopt (median) supermassive BH masses inferred from host galaxy scaling relations \citep[][see Table \ref{tab:redshift_last_det} for their values]{Yao23}.

The injections are done at 0.1 day intervals, from 1 yr after discovery to the last ZTF observation, and the same search is repeated for each injection made. By requiring each signal model to not be recovered by the search with $>5\sigma$ significance in all ZTF bands, we set the constraint on the rate of the signal. We assume the flares occur according to a Poisson process, a model-agnostic simplification which could be potentially improved in future works by employing detailed model predictions of the flare pattern.

\begin{figure}
    \centering
    \includegraphics[width=\linewidth]{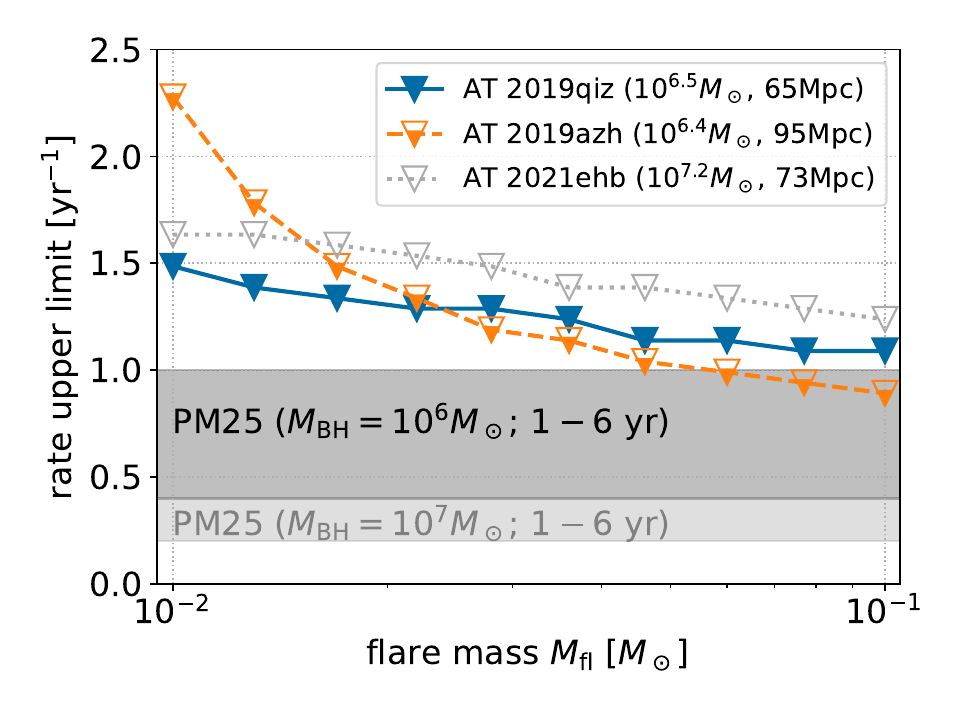}
    \caption{90\% confidence upper limits on the flare event rate as a function of flare mass $M_{\rm fl}$, for the three TDEs, AT 2019qiz, AT 2019azh, and AT 2021ehb (parentheses denote their estimated BH masses and distances). Shaded regions show the theoretical predictions from \cite{Piro25}, for two BH masses ($10^6$, $10^7~M_\odot$) and over 1-6 years since disruption.}
    \label{fig:rate_constraints}
\end{figure}
Figure \ref{fig:rate_constraints} shows the 90\% confidence upper limit on the event rate as a function of the flare mass $M_{\rm fl}$. We overall find a constraint of $\lesssim 1$--$2$ yr$^{-1}$ for each of the three TDEs, stronger for higher $M_{\rm fl}$. For the two closer TDEs AT 2019qiz (65 Mpc) and 2021ehb (73 Mpc), the dependence on $M_{\rm fl}$ is rather weak. This is because the flare is mostly detectable for the entire range of $M_{\rm fl}$ when observed around peak. The limit is primarily set by seasonal gaps, with a secondary effect of larger $M_{\rm fl}$ leading to transients with slightly longer durations. Thus for these two events, we expect the limits to be robust to variations of the spectral energy distribution (SED) for the injections. For the farthest AT 2019azh (95 Mpc), the dependence on $M_{\rm fl}$ becomes stronger, with weaker limits at low $M_{\rm fl}$ due to limitation, in sensitivity. Therefore the limits from this event (and all other events that are farther away) would be sensitive to variations of the SED from our assumptions, especially for lower $M_{\rm fl}$.

In Figure \ref{fig:rate_constraints} we also show in shaded regions the theoretical expectations from the disk instability model of \cite{Piro25}, for two supermassive BH masses of $10^6, 10^7~M_\odot$ close to the measured BH masses of the three events. At $1$--$6$ years from TDE that these events cover, \cite{Piro25} predict flares occurring 2--5 (1--2) times for $M_{\rm BH}=10^6~M_\odot$ ($10^7~M_\odot$). The flare rate is controlled by the rate of the disk's mass gain by fallback accretion, as well as the disk's viscous evolution that controls the cycling behavior between low and high states. While there are no simple functional forms for predicting the occurrence rate, we expect smaller flare rates when the global accretion rate is smaller, e.g. for (i) a smaller $M_*$, (ii) a larger $M_{\rm BH}$, and (iii) a smaller viscosity parameter $\alpha$. The flare rate also decreases with time from disruption, as the mass fallback rate drops and the disk takes longer time to accumulate enough mass for the instability to occur. For their fiducial model ($\alpha=0.1, M_*=M_\odot$) with a BH mass of $10^6~M_\odot$ ($10^7~M_\odot$), the flares completely cease at $\approx 5$ (3) years from TDE.

Taking into account the range in the theoretically expected rates, our constraints of $\lesssim 1$--$2$ yr$^{-1}$ do not immediately rule out this model, but are quite marginal. We expect the search to become more sensitive in the future, with longer baselines and higher sensitivities to be realized for more TDEs.

\section{Conclusion}
\label{sec:conclusion}
We have studied the transient UV and optical emission from  accretion disks formed in TDEs, under the picture suggested from recent models that they undergo late-time thermal instabilities. Such instabilities may also explain the emergence of late-time radio emission, seen in a significant fraction of TDEs \citep[e.g.,][]{Wu25}.

We find that the emission from such disk instabilities is expected to be a very hot (a few -- 10 $\times 10^4$ K), luminous transient with luminosities of $10^{42}$--$10^{43}$ erg s$^{-1}$ in the near-UV. The near-UV magnitude peaks between $-17$ and $-19$ mag, which is a promising target for ULTRASAT for TDEs out to redshift $\approx 0.1$. The optical emission is expected to be an order of magnitude fainter, but may still be well constraining in the era of future high-cadence surveys like LSST and Argus Array.

We finally conducted a search for these transient flares in the sample of 28 nearby TDEs (redshift $<0.1$) using optical data from ZTF. We did not find any promising candidates from late-time photometric detections. The non-detection led to upper limits in their event rates of $1$--$2$ yr$^{-1}$, for each of the three most constraining TDEs.

We conclude with suggestions for potential improvements to theoretical predictions and observational searches. First, the disk modeling could be improved by extending the one-zone model \citep[e.g.,][]{Lu22,Piro25} to a more realistic 1D/2D model, and/or including the effects of delayed circularization over a timescale of months to years as suggested in recent numerical studies. This could provide more robust predictions for both the timing and the mass of the flares. Second, the greybody assumption for the spectra could be improved by detailed radiative transfer modeling of the outflow, which could impact more on the predictions for optical that are far from the spectral peak. Finally the search can be improved in several ways as more sensitive surveys are realized, such as combining information of multiple filters and applying more physically informed priors on the flares for significance estimates and rate constraints.

\begin{acknowledgments}
We thank Kate Alexander and Noah Franz for discussions on OTTER data, Xiaoshan Huang, Brenna Mockler, David Sand and members of Edo Berger's group for discussions, and Assaf Horesh for comments that motivated this study. We further thank the anonymous referee for the constructive comments that improved this manuscript. D. T. is supported by Harvard University through the Institute for Theory and Computation Fellowship. The Villar Astro Time Lab acknowledges support through the David and Lucile Packard Foundation, the Research Corporation for Scientific Advancement (through a Cottrell Fellowship), the National Science Foundation under AST-2433718, AST-2407922 and AST-2406110, as well as an Aramont Fellowship for Emerging Science Research. This work is supported by the National Science Foundation under Cooperative Agreement PHY-2019786 (the NSF AI Institute for Artificial Intelligence and Fundamental Interactions).  

Code for reproducing the results are publicly available in \url{https://github.com/DTsuna/TDEflares}.
\end{acknowledgments}

\appendix
\twocolumngrid

\section{Generarization to a Time-Varying accretion flare}
Our main text has assumed a top-hat temporal profile of the accretion flare, with constant $\dot{M}_w$ and $L_{\rm acc}$ over a duration $t_{\rm fl}$, to make the problem analytically solvable. Here we generalize this for a time-dependent outflow, with wind outflow rate of $\dot{M}_w(t)$ ($0<t<t_{\rm fl}$) and an accretion power of $L_{\rm acc}(t)=GM_{\rm BH}\dot{M}_w/r_{\rm in}$. We fix the inner radius $r_{\rm in}$ and wind velocity $v_w\approx \sqrt{GM_{\rm BH}/r_{\rm in}}$ to be constants in time, setting $r_{\rm in}\approx r_d$ that is consistent with the top-hat case for our parameters of interest (see Section \ref{sec:model}).

\begin{figure*}
    \centering
\includegraphics[width=\linewidth]{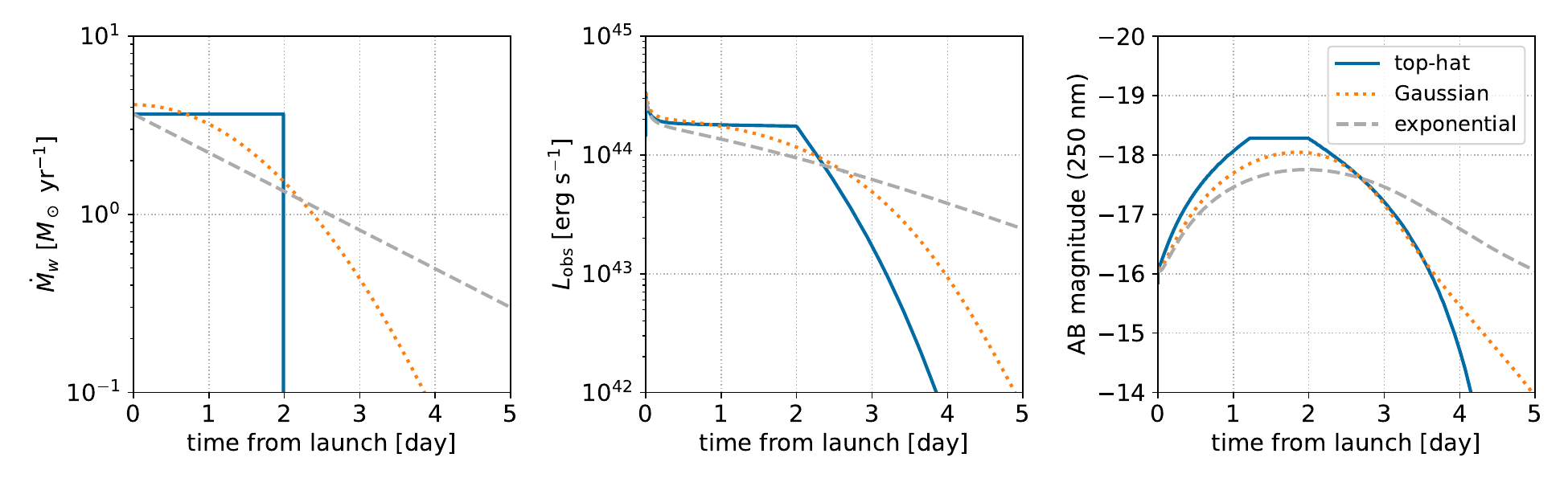}
    \caption{Light curve predictions when the outflow temporal profile is varied from a simple top-hat in the main text. Gaussian and exponential forms are described in equations (\ref{sec:Mdot_exp}) and (\ref{sec:Mdot_Gaus}). The input parameters are the same as those in Figure \ref{fig:single_model}: $M_{\rm BH}=10^6M_\odot$, $r_d=300R_g$, $M_{\rm fl}=0.02~M_\odot$, and $t_{\rm fl}=2$ days.}
    \label{fig:Mdot_variation}
\end{figure*}

The approach follows that in the main text, of splitting the wind into multiple shells launched at times $\tilde{t}~(0<\tilde{t}<t_{\rm fl})$ over a duration $\Delta t$, and calculating the luminosity radiated by the shell at time $t=\tilde{t}+t'$. We similarly solve for the evolution of the shell's internal energy $E_{\rm int}(t')$
\begin{eqnarray}
    \frac{dE_{\rm int}}{dt'} = -\frac{2E_{\rm int}}{3(t'+r_{\rm in}/v_w)} - \frac{E_{\rm int}}{t_{\rm diff}},
    \label{eq:Eint_eq_timedep}
\end{eqnarray}
for an initial value of $E_{\rm int, 0}=L_{\rm acc}(\tilde{t})\Delta t/2$. The diffusion time $t_{\rm diff}$, evaluated by integrating material ahead of the shell at radius $(r_{\rm in}+v_wt')$, is now replaced by that of a time-dependent outflow rate
\begin{eqnarray}
    t_{\rm diff}(t') &\approx& \frac{[r_w-(r_{\rm in}+v_wt')](r_{\rm in}+v_wt')}{r_w}\nonumber \\
    &&\times \frac{1}{c} \int_{r_{\rm in}+v_wt'}^{r_w}\frac{\kappa\dot{M}_{w,0}(r)}{4\pi r^2 v_w}dr  
    \label{eq:t_diff_timedep}
\end{eqnarray}
where $r_w=r_{\rm in}+v_w(\tilde{t}+t')$ is the outer edge of the wind, and $\dot{M}_{w,0}(r)$ is the outflow rate $\dot{M}_w$ at time $(\tilde{t}+t') - (r-r_{\rm in})/v_w$ when the wind material at radius $r$ was launched. Note that for a top-hat temporal profile in the main text, $\dot{M}_{w,0}$ is independent of $r$ within where the wind exists, and equation (\ref{eq:t_diff_timedep}) reduces to equation (\ref{eq:t_diff}), making equation (\ref{eq:Eint_eq_timedep}) analytically solvable.

For an arbitrary form of $\dot{M}_w(t)$, the equation for internal energy (\ref{eq:Eint_eq_timedep}) could be numerically solved, which gives $E_{\rm int}(t')$ and the radiated luminosity $L_{\rm rad}(t')=E_{\rm int}(t')/t_{\rm diff}(t')$. Then the observed luminosity $L_{\rm obs}$ is given by integration over all the shells like done in equation (\ref{eq:L_rad}),
\begin{eqnarray}
    L_{\rm obs}(t) = \int_0^t d\tilde{t}~ \frac{L_{\rm rad}(t'=t-\tilde{t})}{\Delta t}.
\end{eqnarray}
Note that the final $L_{\rm obs}$ is independent of $\Delta t$, as $L_{\rm rad}$ scales with $\Delta t$ and its dependence cancels out.

The temperature of the emission can be similarly solved as done in Section \ref{sec:temperature}, with a density profile at time $t$ now given by
\begin{eqnarray}
    \rho(r,t) = \frac{\dot{M}_{w,0}(r)}{4\pi r^2 v_w}~ (r_{\rm in}<r < r_w),
\end{eqnarray}
where again $\dot{M}_{w,0}$ is the outflow rate at the launching time $t-(r-r_{\rm in})/v_w$. The trapping radius is now defined by finding the radius where the dynamical and diffusion times are equal \citep[Section 3.1 of][]{Piro20}
\begin{eqnarray}
    \frac{r_{\rm tr}-r_{\rm in}}{v_w} = \frac{(r_w-r_{\rm tr})r_{\rm tr}}{r_w} \frac{1}{c} \int_{r_{\rm tr}}^{r_w}\frac{\kappa\dot{M}_{w,0}(r)}{4\pi r^2 v_w}dr ,
\end{eqnarray}
which converges to equation (\ref{eq:r_trap}) for the case of a top-hat temporal profile of $\dot{M}_w$ in the limit of $r_w\gg r_{\rm tr}$. The rest of the formulation for deriving the temperature is equivalent to Section \ref{sec:temperature}.

As a demonstration, we adopt two cases with smoother temporal profiles of $\dot{M}_w(t)$ than a top-hat, an exponential and Gaussian form. We normalize them such that they have a total flare mass of $M_{\rm fl}$ for a characteristic timescale of $t_{\rm fl}$,
\begin{eqnarray}
    \dot{M}_w(t) &=& \frac{M_{\rm fl}}{t_{\rm fl}}\exp\left(-t/t_{\rm fl}\right) \ ({\rm Exponential}), \label{sec:Mdot_exp}\\
    \dot{M}_w(t) &=& \frac{2}{\sqrt{\pi}}\frac{M_{\rm fl}}{t_{\rm fl}}\exp\left[-(t/t_{\rm fl})^{2}\right]\ ({\rm Gaussian}). \label{sec:Mdot_Gaus}
\end{eqnarray}
The corresponding $\dot{M}_w(t)$ and the resulting light curves are shown in Figure \ref{fig:Mdot_variation}. As seen from the left panel, the Gaussian form is close to a top-hat, whereas the exponential has a much smoother tail (linear in a log-linear plot). The late-time luminosity is thus also different, with the exponential having a slower decline than the Gaussian or top-hat models.

The right panel compares the UV light curves at 250 nm. For a Gaussian temporal profile, the light curve is very similar to a top-hat one, giving nearly unchanged predictions for the peak magnitude and duration. The exponential model predicts a prolonged tail in the UV light curve, which lessens the peak magnitude by $\approx 0.5$ mag and increases the duration by factor $\sim 1.5$. We conclude that varying the accretion rate evolution, while keeping the same total mass and characteristic timescale, would not change the light curve properties by more than a factor $\sim 2$.

\section{TDE Sample}
\label{sec:TDE_samples}
We summarize our sample of 28 low-redshift TDEs analyzed in this work (Section \ref{sec:search}). The information for these TDEs and their host-subtracted photometric data were obtained from the literature, via the OTTER database\footnote{\url{https://otter.idies.jhu.edu}} \citep{Franz25} as of December 2025. 

Table \ref{tab:redshift_last_det} shows the redshift, last ZTF detection from discovery, published mass estimates of the supermassive BH, and references for the photometric data and the quoted BH mass. Figure \ref{fig:ZTF_photometry} shows the ZTF light curves of each event, comparing with the expected flare magnitude discussed in Section \ref{sec:prospect}.

\begin{deluxetable*}{ccccc}
\tablecaption{Redshift ($z$), luminosity distance ($D_L$), epoch of last ZTF detection from discovery, and the supermassive BH mass estimates (obtained by host galaxy scaling relation, or light curve modeling if not available), for the 28 TDEs analyzed in this work. $D_L$ is calculated assuming a flat $\Lambda$CDM cosmology with parameters $H_0=70\ {\rm km\ s^{-1}\ Mpc^{-1}}, \Omega_m=0.3$ and $\Omega_\Lambda=0.7$.\label{tab:redshift_last_det}}
\tablewidth{0pt}
\tablehead{
\colhead{Event} & \colhead{ $z$ ($D_L$ [Mpc])} & \colhead{Last detection [yr]} & \colhead{$\log_{10} (M_{\rm BH} [M_\odot])$} & \colhead{References}
}
\startdata
        2018hco & 0.0880 (401) & 6.33  & $6.5\pm 0.3$ & a, b, c\\ 
        2018lna & 0.0910 (416) & 5.83 & $5.2\pm 0.4$ & a, b, d \\ \hline 
        2019azh & 0.0220 (95) & 5.94  & $6.4\pm 0.3$ & a, b, d, e, f, g\\ 
        2019dsg & 0.0512 (227) & 5.39 & $6.9\pm 0.3$ & a, b, d\\ 
        2019ehz & 0.0740 (334) & 2.05 & $5.7\pm 0.6$ & a, b, d\\ 
        2019eve & 0.0813 (369) & 2.98 & $5.1\pm 0.4$ & a, b, d\\ 
        2019qiz & 0.0151 (65) & 5.34 & $6.5\pm 0.3$ & a, b, d, h\\ \hline 
        2020ksf & 0.0942 (431) & 2.59 & $6.1\pm 0.4$ & b, i \\
        2020mbq & 0.0930 (426) & 3.23 & $\approx 6.08$ & b, j\\ 
        2020mot & 0.0700 (315) & 4.60 & $6.7\pm 0.3$ & b, d, j\\ 
        2020neh & 0.0620 (278) & 3.81 & $5.4\pm 0.5$ & b, d, k\\
        2020ocn & 0.0705 (318) & 3.06 & $6.4\pm 0.6$ & b, j, l\\
        2020vdq & 0.0440 (194) & 2.74 & $5.6\pm 0.4$ & b, d, m\\
        2020zso & 0.0565 (252) & 2.70 & $6.2\pm 0.3$ & b, j, n\\ \hline
        2021ehb & 0.0170 (73) & 3.89 & $7.2\pm 0.3$ & b, d, o\\ 
        2021mhg & 0.0730 (329) & 2.35 & $6.1\pm 0.4$ & b, d\\
        2021nwa & 0.0470 (208) & 3.17 & $7.2\pm 0.3$ & b, d, p \\
        2021sdu & 0.0590 (264) & 2.91 & $6.7\pm 0.3$ & b, d\\ \hline
        2022bdw & 0.0378 (166) & 2.62 & $6.2\pm 0.5$ & b, p\\
        2022dbl & 0.0284 (124) & 6.10 & $6.4\pm 0.3$ & b, q \\
        2022dyt & 0.0720 (325) & 1.13 & $6.3\pm 0.3$ & b, r\\ 
        2022exr & 0.0955 (438) & 2.51 & $5.4\pm 0.3$ & b, r\\
        2022gri & 0.0280 (122) & 2.83 & $6.2\pm 0.3$ & b, s\\ 
        2022lri & 0.0328 (143) & 1.39 & $5.0\pm 0.5$ & b, t\\
        2022pna & 0.0950 (435) & 1.45 & $6.7\pm 0.4$ & b, r \\
        2022wtn & 0.0491 (218) & 2.25 & $\approx 6.1$ & b, u\\ \hline 
        2023clx & 0.0111 (47) & 1.27 & $\approx 6$ & b, v\\
        2023cvb & 0.0710 (320) & 1.71 & $7.5\pm 0.2$  & b, r\\ 
\enddata{}
\tablerefs{a: \cite{2021ApJ...908....4V}, b: \cite{2024MNRAS.527.2452M}, c: \cite{Alexander25}, d: \cite{Yao23}, e: \cite{2021MNRAS.500.1673H}, f: \cite{2022ApJ...925...67L}, g: \cite{2024ApJ...969..104F}, h: \cite{2020MNRAS.499..482N}, i: \cite{Wevers24}, j: \cite{Hammerstein23}, k: \cite{2022NatAs...6.1452A}, l: \cite{Pasham24}, m: \cite{2025ApJ...982..163S}, n: \cite{Wevers22}, o: \cite{2022ApJ...937....8Y}, p: \cite{Mummery25}, q: \cite{Lin24}, r: \cite{Mummery26}, s: \cite{Zhu25},  t: \cite{2024ApJ...976...34Y}, u: \cite{Onori25}, v: \cite{Charalampopoulos24}}
\tablecomments{For AT 2022dbl there is an observation in 2018 labeled as ``detection" by OTTER, which could be an image subtraction artifact \citep{Makrygianni25}. While we include this as discovery, whether including this as detection or not does not influence our conclusions.}
\end{deluxetable*}
\clearpage

\begin{figure*}
    \centering
    \includegraphics[width=\linewidth]{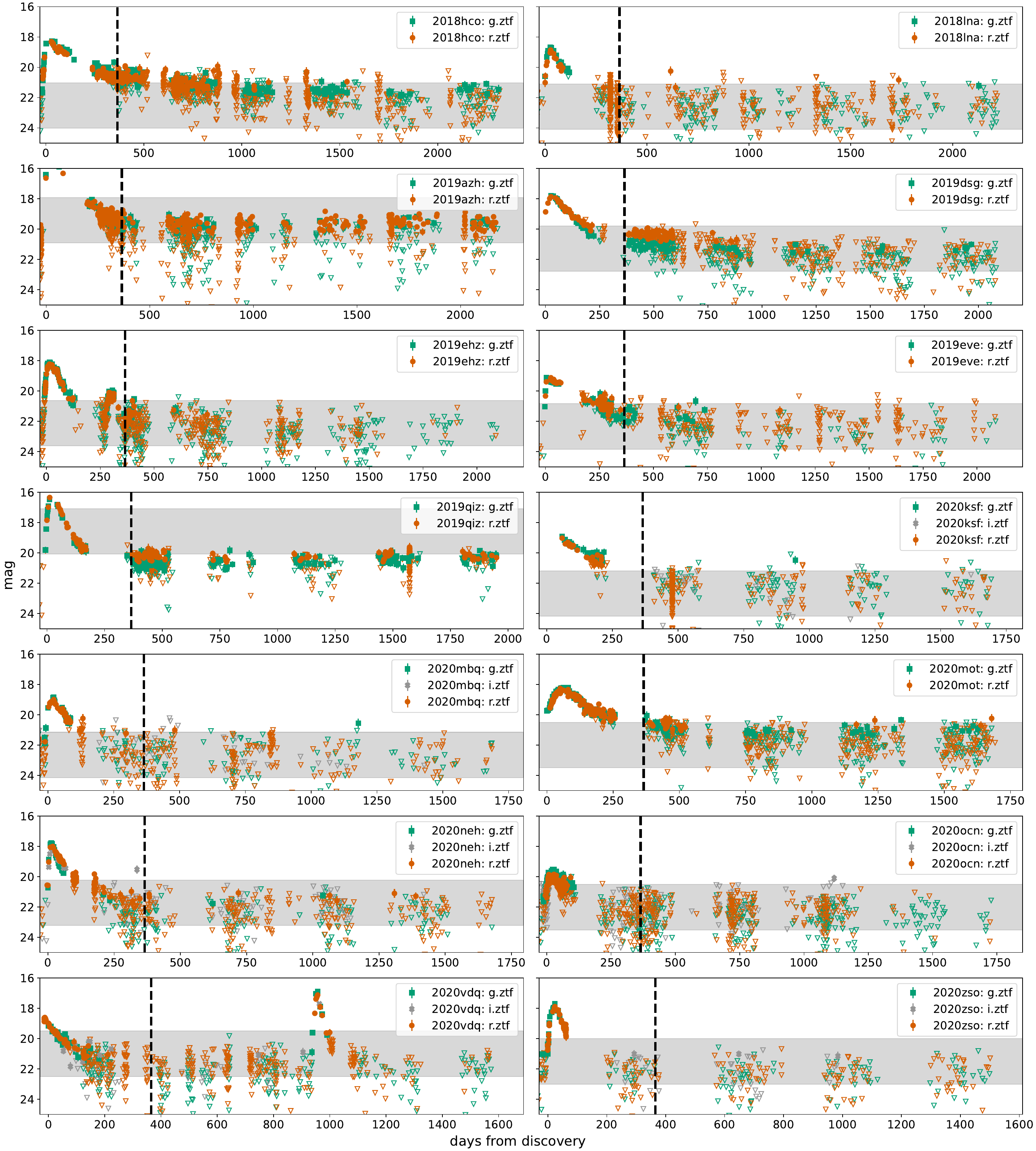}
    \caption{Host-corrected ZTF light curves of 28 TDEs selected from OTTER. Dashed lines show 1 year from discovery, and we search for flares beyond this epoch. Filled circles denote detections (S/N$>3$), and open triangles denote upper limits (S/N$<3$). Shaded regions show the  flare magnitude for an expected absolute  magnitude of (-14) -- (-17) mag (see Section \ref{sec:prospect}).}
    \label{fig:ZTF_photometry}
\end{figure*}
\begin{figure*}
    \ContinuedFloat
    \centering
    \includegraphics[width=\linewidth]{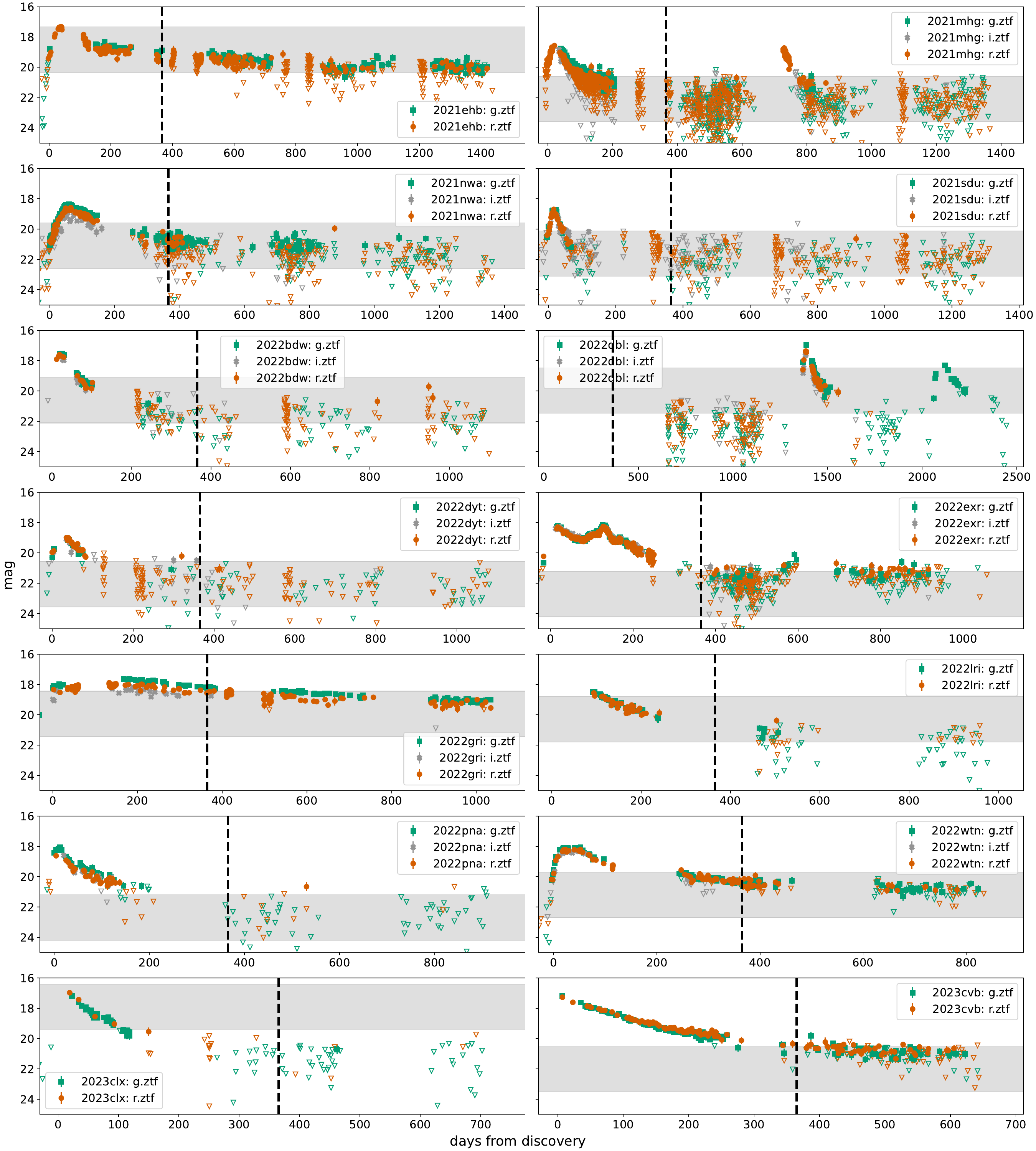}
    \caption{(Continued)}
\end{figure*}

\bibliography{references}
\bibliographystyle{aasjournal}

\end{document}